\documentclass[pdflatex,sn-apa]{sn-jnl}

\usepackage{graphicx}%
\usepackage{multirow}%
\usepackage{amsmath,amssymb,amsfonts}%
\usepackage{amsthm}%
\usepackage{mathrsfs}%
\usepackage[title]{appendix}%
\usepackage{xcolor}%
\usepackage{textcomp}%
\usepackage{manyfoot}%
\usepackage{booktabs}%
\usepackage{algorithm}%
\usepackage{algorithmicx}%
\usepackage{algpseudocode}%
\usepackage{listings}%
\usepackage{microtype}
\usepackage{natbib}
\usepackage{geometry}
\usepackage{setspace}
\usepackage[autostyle]{csquotes}
\usepackage{hyperref}
\usepackage[dont-mess-around]{fnpct} 
\usepackage{todonotes}

\hypersetup{
	colorlinks=true,
	linkcolor=blue,
	citecolor=blue,
	urlcolor=blue
}

\theoremstyle{thmstyleone}%
\theoremstyle{thmstyletwo}%

\theoremstyle{thmstylethree}%

\graphicspath{{images/}}

\newcommand{\argmax}{\operatornamewithlimits{argmax}}

\begin{document}

\title[DAA Improves Peer Review Process]{Deferred Acceptance Algorithm Improves Peer Review Process}


\author*[1]{\fnm{Christoph} \sur{Bartneck}}\email{chirstoph.bartneck@canterbury.ac.nz}
\equalcont{These authors contributed equally to this work.}

\author[2]{\fnm{Richard} \sur{Watt}}\email{richard.watt@canterbury.ac.nz}
\equalcont{These authors contributed equally to this work.}

\author[1]{\fnm{Etienne} \sur{Borde}}\email{etienne.borde@canterbury.ac.nz}
\equalcont{These authors contributed equally to this work.}

\author[2]{\fnm{Pattara} \sur{Klinpibul}}\email{pattara.klinpibul@pg.canterbury.ac.nz}
\equalcont{These authors contributed equally to this work.}

\affil*[1]{\orgdiv{Computer Science and Software Engineering}, \orgname{University of Canterbury}, \orgaddress{\street{Private Bag 4800}, \city{Christchurch}, \postcode{8041}, \country{New Zealand}}}

\affil[2]{\orgdiv{Department of Economics and Finance}, \orgname{University of Canterbury}, \orgaddress{\street{Private Bag 4800}, \city{Christchurch}, \postcode{8041}, \country{New Zealand}}}



\abstract{The peer review process is essential to the success of science, but it also delays publications and absorbs considerable effort. Journals find it increasingly difficult to recruit competent reviewers. This study presents the results of agent-based simulation that models the current peer review process. We compared it to the simulation of a new peer review process that uses the Deferred Acceptance Algorithm (DAA) to match papers to journals. The matches are just as good while dramatically reducing the required number of reviews and delays. The results show that it is possible for the scientific community to significantly optimise the peer review process.}





\maketitle

\section{Introduction}
The peer review process is resource intensive but essential to science. Even with its widely discussed shortcomings it remains the standard form of quality control for research \citep{RN1578,RN1579}. One of the main problems with the process is the publication delay which consists of the time from submission to acceptance and furthermore from acceptance to publication. Depending on the discipline, it can take on average up to a year for a paper to become accepted \citep{solomon2013}. Papers may go through several iterations before they get accepted and editing the manuscript does take time, but the majority of the delays can most likely be attributed to the time reviewers take to complete their review.

A second main problem with the peer review process is that it is inherently resource intensive. Assuming that each paper gets reviewed on average by three reviewers, then the total number of reviews necessary is three times the number of papers submitted. Given the ever-increasing productivity of science \citep{olesen2010}, the number of reviews required becomes increasingly difficult to acquire. Editors struggle to find reviewers willing to review yet another paper. A contributing factor is that all of this work is done without any financial gain for the editors or reviewers. The system relies on the good will and commitment of the researchers.

In recent years, some innovations to the peer review process have been tested, such as motivating reviewers by acknowledging their contributions \citep{sandeep2016} while maintaining the ``blindness'' of the process. But also, more open approaches have been tried, such as the Open Peer Commentary. In \citet{peters1982} paper on an empirical study on the reliability of the peer review process, the open peer commentary takes up ten times the number of pages as the original target article. Their landmark study also highlights another challenge. Performing empirical studies on the peer review process can be perceived as a form of whistle-blowing, and hence the authors may suffer from the consequences. Jeffrey Beall at the University of Colorado maintained a list of potentially predatory journals and received several threats of lawsuits, in particular from the OMICS group \citep{chappel2013}. The OMICS group was sentenced in 2019 to pay 50 million USD in damages for deceptive practices. One of the conferences operated by OMICS did, for example, accept a nonsensical paper that was written by the auto-complete function of iOS \citep{hunt2016}. \citet{bohannon2013} conducted a sting on open access publishers and the OMICS group again was featured on this black list of publishers who do not use a respectable peer review process.

Jeffrey Beall's list was taken offline, and it has been reported that Beall ``was forced to shut down the blog due to threats \& politics'' \citep{chawla2017}. This shows that experimenting with the peer review system can entail negative consequences. 

But besides the political and legal battles around the publishing process, it also requires considerable resources to run experiments on an operational review process. One challenge is that to run a complete experiment of the quality and reliability of the peer review process, it is necessary to also consider the rejected papers \citep{bartneck2017}. Normally, only the accepted papers are publicly known, and hence the collaboration of the editorial team is necessary. In 2015, the editorial team of the Neural Information Processing Systems (NIPS) conference\footnote{\url{https://nips.cc/}} performed an experiment in which 166 (10\%) of the papers submitted to the conference were put through the peer review process twice \citep{price2014,langford2015}. \citet{price2014} concluded that ``most papers at NIPS would be rejected if one reran the conference review process''. The NIPS conference attracted 1838 paper submissions in 2015, which resulted in 10,625 reviews by 1524 reviewers. Managing the review process for such a large conference is complex, but it also enables researchers to investigate the review process itself, although at a considerable price. 

Running such initial empirical studies is certainly valuable, but changing the policies, processes and traditions of the scientific remains difficult. Systematic manipulations of the factors influencing the process remains unpractical. Agent-based social simulation offers us the opportunity to test new approaches to the scientific publishing process. \citet{guang2005} conducted a simulation to better understand the publication delay and they concluded that the number of submissions to a journal dramatically influences the publication delay.

In this study, we investigated the degree to which the Deferred Acceptance Algorithm (DAA), for which Lloyd Shapley and Alvin Roth received the 2012 Nobel Prize in economics, can be used to match scientific manuscript submissions to scientific journals. \citet{gale1962} already showed that their algorithm produces stable matches when both sets are equally sized. They successfully applied their algorithm to a variety of problems over the years, such as the matching of students and schools or the matching of organ donors to recipients \citep{roth2015}.

The scientific publishing process is also a form of match-making in which money traditionally is not used to regulate the market. At least not on the level of the individual researcher and journal. Authors submit their papers to journals or conferences, and neither the author nor the journal pays any money at this point. There is, however, a market regulated by money between the journal publishers and the university libraries, but we shall not investigate this market here since it is normally of no direct relevance to an individual researcher who has to select where to submit his/her paper. Some universities, however, started to have incentives for researchers to publish in reputable journals, such as those listed in Scopus. 

The recent rise of predatory journals and conferences that publish any paper for a fee may disrupt this process. It has already been argued that any paper that gets written gets published eventually \citep{bartneck2010}. The predatory journals only strengthen this point of view since also complete nonsense papers get published \citep{sokal1996,safi2015,lindsey2017,hunt2016}. The market might also slightly change due to new legislation in the EU and USA, which forces authors funded through certain funding agencies to publish their papers in an open access venue \citep{RN1592}. All traditional publishers already offer open access options, although at a considerable price.

This study focuses on what contribution DAA can make to better match manuscripts with journals. Currently, authors submit their paper to only one venue that reviews the submission. In case the reviews are unfavourable, then the editor rejects the paper and the author submits the manuscript again to either the same journal or a completely new journal. The author will hopefully have used the comments of the reviewers to improve the paper, but this cannot be guaranteed. This serial process continues until the manuscript is finally matched to a journal. The resulting delays are a considerable problem \citep{solomon2013}.

The DAA can improve this process by matching all submitted papers to all journals in one step. Authors submit their manuscripts to a central warehouse and rank the journals according to their preference. The submitted manuscripts are treated as a commodity; the central warehouse organises the evaluation of every submitted paper, and all the journals receive access to the reviews, similar to how commodities are classified these days \citep{roth2015}. The journals then submit their preference list to the central warehouse and the DAA matches manuscripts to journals. This scenario has the potential advantage of minimising both the number of required reviews and the review delay.

To compare these two DAA scenarios with the status quo, we created and executed agent-based simulations using the Python-based framework Mesa \citep{masad2015mesa,kazil2020utilizing,ter2025mesa}. We measured the resulting workload for the reviewers and the review delay as efficiency indicators, and we measured the matching fit and satisfaction of the researchers as quality indicators. The remaining structure of this paper is as follows: Section~\ref{sect: submission model} introduces a microeconomic model of the status quo submission process; Section~\ref{sect: simulation model} outlines the overall simulation framework; Section~\ref{sect: status quo} applies the status quo submission model to the simulation; Section~\ref{sect: DAA} applies the DAA matching algorithm to the simulation; Section~\ref{sect: parameters} discusses the chosen simulation parameters; Section~\ref{sect: results} presents the simulation results and discusses its implications; and, lastly, Section~\ref{sect: conclusion} concludes.

\section{A simple microeconomic model for journal submissions} \label{sect: submission model}

We start with the presentation of a microeconomic approach to model the traditional paper submission process. Consider a situation in which an author has written an article, and they are searching for a journal to publish it. The author will set up a ranking of the available journals, and begin by submitting to the highest one on the ranking. If the article is accepted, then the game stops. If the article is rejected, then the author submits to the next highest journal on their ranking. This goes on until either the paper is accepted or all of the journals on the ranking list have rejected the paper.

This situation has features that are, formally, similar to a closed-bid first-price auction, and other features that are similar to a sequential search problem. Essentially, the author is the seller, and the article is the good being sold. The journals are the buyers (bidders). Each journal has a known \enquote{value} or payoff for the author (the value of having the article accepted at that journal), and each has a (subjectively) known probability of accepting the article. Each journal, therefore, presents to the author a \enquote{prospect}, which is a two-dimensional vector $\pi=(p,v)$, where $p$ is the probability of acceptance, and $v$ is the value to the author of having their paper accepted. We will assume that for journal $k$ with probability of acceptance $p_k$, then, if the article is accepted, the payoff is $v_k=1/p_k$. Furthermore, assume that there is a continuum of journals that are ordered on $p\in(0,1]$. So, the very lowest-ranked journal gives acceptance for sure, with a payoff equal to 1, while the highest-ranked journal has a probability arbitrarily close to 0, and a valuation that approaches infinity.

First, let us consider a one-shot game, where the author has only one attempt at a journal submission. If the chosen journal rejects, the paper is directly published by the lowest-ranked journal ($p=1$). Assume that having the paper reviewed by the journal takes one \enquote{period} of time, so if the journal rejects, then the author passes into a new time-period, and correspondingly, utility is discounted by a factor $\beta\in[0,1]$.%
\footnote{$\beta$ quantifies how much less a person would value something in the future; it represents the patience of an individual, where the edge case of $\beta=1$ would mean infinite patience.}

Defining $u(x)$ as the utility function, the expected utility of choosing a journal with probability $p$ is
\begin{equation} \label{eq: expected utility}
    Eu(p)=pu\left(\frac{1}{p}\right) + (1-p)\beta u(1),
\end{equation}
such that,
\begin{equation} \label{eq: oneshot game}
    p^*=\argmax_p Eu(p).
\end{equation}
The second-order condition (SOC) is met if we assume risk aversion ($u''<0$). To that end, we assume that the utility function satisfies constant relative risk aversion (CRRA),%
\footnote{This is a very common assumption in simulation experiments in economics, since it captures in a very simple and intuitive manner the effect of potential changes in risk aversion.}
\begin{equation} \label{eq: utility}
    u(x)=\frac{x^{1-r}-1}{1-r},
\end{equation}
where $r$ is the constant level of relative risk aversion%
\footnote{This is also sometimes referred to as the elasticity of marginal utility of consumption (EMU).}
, which is not equal to $1$---the limit of $u(x)$ as $r\rightarrow 1$ is $\ln{(x)}$. Substituting \eqref{eq: utility} into \eqref{eq: oneshot game} yields
\begin{equation}
    p^*(r)=r^{\frac{1}{1-r}}.
\end{equation}
Note that as $r\rightarrow1$ we get $p^*=1/e\cong0.3679$. More importantly, the optimal probability in the one-shot game is independent of $\beta$. Also note that $p^*$ is an increasing function of $r$; the intuition is that as relative risk aversion increases, the author should choose to submit to a journal with a higher probability of accepting the paper, as this is a less risky choice.

\subsection{Multiple submission periods} \label{subsect: multiple submission}

Now, we extend the prior example to multiple submission periods. We set up a recursive sort of market, in which rejection by the most favoured journal leads to submission to a less favoured journal, but still not the default one. Thus, the decision is to find the starting point journal, say $p_1^*$, that then sets in place the potential sequence of submissions. For example, say there is only time to try at two journals with a probability less than 1 (and then the default journal), then the author's problem is
\begin{equation} \label{eq: twoshot game}
    p_1^*,p_2^*=\argmax_{p_1,p_2} \, p_1u\left(\frac{1}{p_1}\right)+(1-p_1)
        \beta\underbrace{\left[p_2u\left(\frac{1}{p_2}\right)+(1-p_2)\beta u(1)\right]}
            _{\text{one-shot game}}.
\end{equation}
Notice that the square bracket term in \eqref{eq: twoshot game} is exactly the same as the one-shot game; see \eqref{eq: expected utility} and \eqref{eq: oneshot game}. Thus, the solution to the two-shot game would be easily found by recursively substituting in as $p_2^*$ the one-shot game solution, and then resolving the first stage of the game.

Following this, we can easily work out the optimal sequence of journals for a multi-period game with any number of periods. Say there are $n$ periods in total, including the potential last period in which the author would submit to the journal with $p=1$. Period 1, then, is the first submission, period 2 is the second submission (should period 1 result in rejection), and so on. Consider what the author would do if they arrive at period $i$. The rejections that have happened prior to period $i$ are now irrelevant to the decision regarding $p_i^*$. Let us define
\begin{equation} \label{eq: nshot game}
    U_i^*(p_i)\equiv\argmax p_iu\left(\frac{1}{p_i}\right) + (1-p_i)\beta U^*_{i+1}(p_{i+1}), 
        \quad \text{for } i\in\{1,...,n-1\},
\end{equation}
and $U_n^*(p_n)=u(1)$.

\begin{figure}[h!]
    \centering
    \begin{minipage}{0.5\textwidth}
        \centering
        \includegraphics[width=\linewidth]{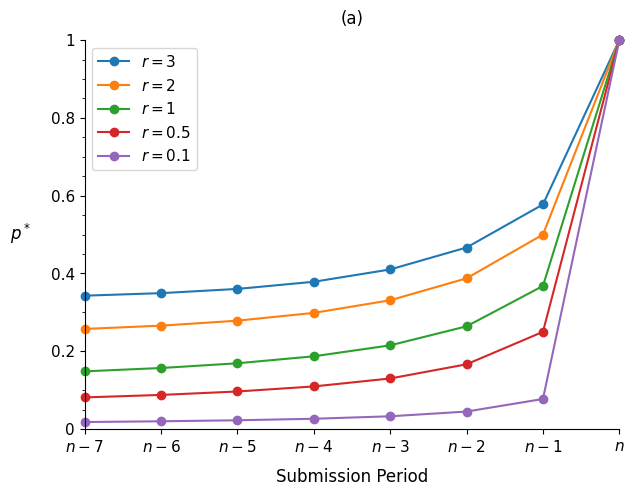}
    \end{minipage}%
    \begin{minipage}{0.5\textwidth}
        \centering
        \includegraphics[width=\linewidth]{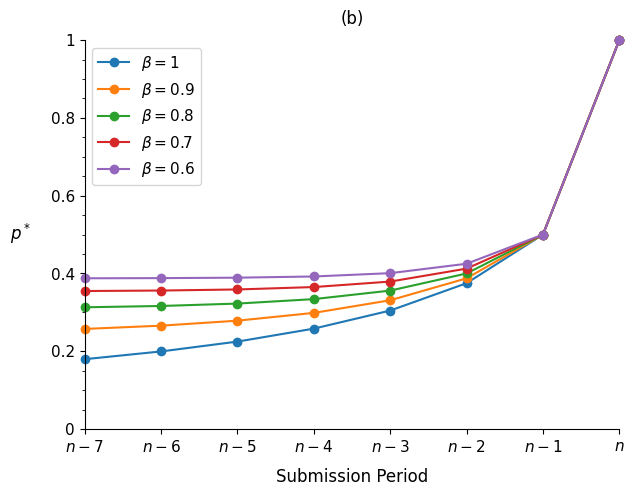}
    \end{minipage}
    \caption{Optimal submission plan for up to $n=8$ submission periods for (a) different levels of $r$ when $\beta=0.9$, and (b) different levels of $\beta$ when $r=2$}
    \label{fig: multistage game}
\end{figure}

The first-order condition for \eqref{eq: nshot game} is
\begin{equation} \label{eq: nshot FOC}
    u\left(\frac{1}{p_i^*}\right)-u'\left(\frac{1}{p_i^*}\right)\frac{1}{p_i^*}=\beta U_{i+1}^*, \quad
        \text{for } i\in\{1,...,n-1\}.
\end{equation}
This establishes the sequence of journals that make up the submission plan, $\mathbf{p^*}=(p_1^*,...,p_{n-1}^*,1)$, which is solved recursively, first calculating $p_{n-1}^*$, then $p_{n-2}^*$, and so on, until finally calculating $p_1^*$. In \eqref{eq: nshot FOC}, since $u''<0$, we can note that $\beta U_{i+1}^*$ is a decreasing function of $1/p_i$. Thus, the greater (resp. smaller) $\beta U_{i+1}^*$ is, the smaller (resp. greater) $1/p_i^*$ is---i.e., the greater (resp. smaller) $p_i^*$ is. Therefore, given $\beta>0$, if $U_i^*>U_{i+1}^*$ for all $i$, then the sequence of probabilities in the optimal submission plan $\mathbf{p^*}$ is increasing, that is, $p_i^*<p_{i+1}^*$ for all $i$.%
\footnote{An interesting but highly unlikely case is $\beta=0$, for which the problem faced is identical in each period, and so the author would choose the same $p_i^*=p^*$ for all $i<n$ (and of course, $p_n^*=1$). Other cases, such as a monotone decreasing sequence or the case in which the author optimally chooses $p^*=1$ in the very first period, are also possible.}
Figure~\ref{fig: multistage game} shows the optimal submission plan for up to 8 submission periods for varying levels of $\beta$ and $r$.

One very interesting point to consider is the optimal number of periods that the author will want to use for submitting the paper. The author should be allowed to choose how many submissions to make before yielding to the journal of universal acceptance. Intuitively, for each period of searching for a journal, the author gains a certain probability of a better payoff but suffers the cost of delaying the payoff in the event that the new search yields a rejection. The optimal number of submissions will balance these two effects at the margin. In short, if there is a period $i\in\{1,...,n\}$ for which $U_i^*(p_i)<\beta U_{i+1}^*(p_{i+1})$, the author will prefer not to use all $n$ submission periods, but if $U_i^*(p_i)>\beta U_{i+1}^*(p_{i+1})$ for all $i$, then the author will benefit from adding more periods in which to submit (i.e., increasing $n$).

\section{Publication simulation model} \label{sect: simulation model}

Our simulation model focuses on an academic discipline consisting of researchers and journal agents. The researchers write and review papers, and the journals publish them. The agents sequentially move each time step. The time step, $\Delta t$, is defined as 1 day, and we use a random scheduler to determine the order of the moves each step. The model will be used to compare the traditional (status quo) publication process with the Deferred Acceptance Algorithm (DAA). 

In our model, researchers ($R$) write papers and review the papers of others. Each researcher $i\in \{1,...,N_R\}$ is assigned a quality, $q_i^R\in (0,1)$, where $N_R$ is the total number of researchers, and a higher quality rating corresponds to a better researcher who writes better papers. We assume that each researcher perceives their own true quality; that is, the perceived quality of a researcher is perfect information. Additionally, as we will see later on, we also assume that the researcher is able to correctly perceive the qualities of their own papers and is thereby immune to the Dunning-Kruger effect. Researchers produce papers at different speeds, and thus, each researcher is assigned a writing productivity, $\eta_i^w$, which represents the number of days it takes the researcher to write a paper. We do not consider the time it takes to conduct the actual research, since this can vary dramatically. Note that we assume that the researchers only work on one paper at a time. Each researcher also has a response time, $\eta^{\mathrm{resp}}_i$, and a reviewing productivity, $\eta^{\mathrm{rev}}_i$, which respectively represent the amount of days it takes the researcher to respond (accept/decline) to a review invitation and the amount of days it takes the researcher to review a paper, once an invitation has been accepted. We simplify and assume that $\eta^{\mathrm{rev}}_i=\eta^{\mathrm{rev}}$ and $\eta^{\mathrm{resp}}_i=\eta^{\mathrm{resp}}$. Lastly, we assume that each reviewer can only review up to three papers at a time.%
\footnote{A limitation is that, in reality, a researcher's capacity to review may also be affected by other workload.}

The researchers produce papers ($P$) with quality $q^P_{i,j}\in(0,1)$ relative to their own quality with some deviation $\varepsilon^{P}_{i,j}$, such that, for paper $j$ and researcher $i$,
\begin{equation}
    q^{P}_{i,j}=q^{R}_i + \varepsilon^{P}_j.
\end{equation}
Unlike researchers, the perceived quality of a paper is imperfect information. We will use a \enquote{hat} to denote perceived information, and we apply \enquote{noise} to generate imperfect information. For each paper that a researcher reviews, they will score the paper based on their perceived quality of the paper, $\hat{q}^P_{i,j}\in(0,1)$. Each researcher evaluates the quality of each paper slightly differently with variation $\varepsilon^{\mathrm{rev}}_{i,j}$, such that, for paper $j$ and reviewer $i$,
\begin{equation} \label{eq: review score}
    \hat{q}^{P}_{i,j}=q^{P}_{j} + \varepsilon^{\mathrm{rev}}_{i,j}.
\end{equation}

The second class of agents in our model are journals ($J$); similarly as for the researchers, in our model, each (standard peer-reviewed) journal $k\in\{1,...,N_J\}$ has a quality $q^J_k\in(0,1)$, where $N_J$ is the total number of peer-reviewed journals, and a higher quality rating corresponds to a better journal. In contrast to the researchers, the quality of a journal is updated each time it accepts a new paper; each journal is given an initial quality $q^{J_0}_k$. The quality of journal $k$  is recalculated as the mean quality of accepted papers, where the initial quality has a weight of 1,000 papers. Let us define $A_k(t)$ as the (current) set of accepted papers for journal $k$, such that
\begin{equation} \label{eq: journal quality updating}
    q^J_k(t) = \frac{1,000q^{J_0}_k+\sum_{j\in A_k}q^P_{j}}{1,000+|A_k(t)|}, \quad k\in\{1,...N^J\}.
\end{equation}
Additionally, We define a special-case journal $k=0$ as an all-acceptance journal, which (instantly) accepts every paper submitted without peer review. Unlike the other journals, its quality is fixed at $q^{J_0}_0=q^J_0=0$. The all-acceptance journal already exists in the form of repositories. The frustrated researcher might, for example, publish the paper in their university's research repository or on their own website. If the author published the paper in a pre-publishing platform, such as Arxiv\footnote{\url{http://arxiv.org/}}, they might just leave it there and not pursue publishing the paper any further. We assume the perceived journal quality also as imperfect information. Researcher $i$ perceives the quality of journal $k$ at a given time $t$ as
\begin{equation}
    \hat{q}^J_{i,k}(t)=q^J_k(t)+\varepsilon^{J}_{i,k,t}.
\end{equation}
Note that we assume the quality of the all-acceptance journal is perfect information ($\hat{q}^J_0=q^J_0=0$).

\section{Status quo publication process} \label{sect: status quo}

We use the microeconomic model for journal submissions (presented in Section~\ref{sect: submission model}) to create our first simulation scenario. The status quo simulation models the currently used conventional peer review process. Researchers write papers and submit them to journals. The journals send the papers to researchers for their review. After receiving all the reviews, the journals make the decision to either accept or reject the paper. This is a slightly simplified process since papers can, of course, go through several iterations of revision before final acceptance, however, we leave this for future research.

\subsection{Researchers' belief system}

\begin{figure}[t!] 
\centering
\includegraphics[scale=0.7]{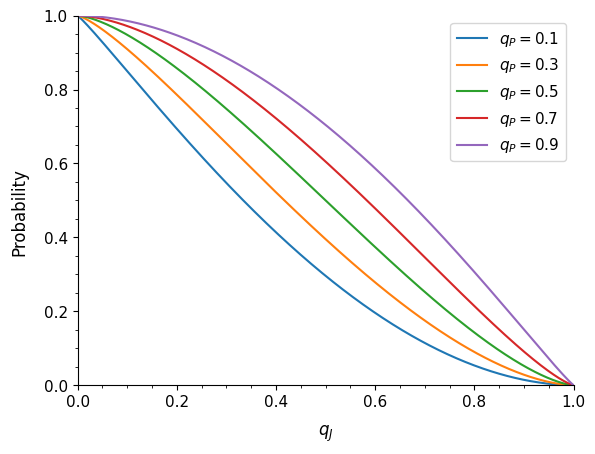}
\caption{Researchers' belief on journal acceptance probability.}
\label{fig: paper prob}
\end{figure}

We first define a universal belief system that governs the behaviour of the researchers. The researchers believe that the likelihood of rejection for a paper of quality $q_P$ for a journal with quality $q_J$ follows a Beta distribution with a probability density function (PDF) of
\begin{align}
    f_{\mathrm{beta}}(q^J;1+q^P,2-q^P)=\bigl[q^J\bigr]^{q^P}\bigl[1-q^J\bigr]^{1-q^P}.
\end{align}
It thus follows that the researcher believes the probability of a paper getting accepted is
\begin{equation}
    p(q^P,q^J) = 1-F_{\mathrm{beta}}(q^J;1+q^P,2-q^P),
\end{equation}
where $F_{\mathrm{beta}}(\cdot)$ is the corresponding cumulative distribution function (CDF) of the Beta distribution---since we assume imperfect information, $\hat{p}(q^P,\hat{q}^J)$ is the corresponding perceived probability. Note that, for the all-acceptance journal, we assume that it is perfect information that the acceptance probability is 1 ($\hat{p}_0=p_0=1$).  Figure~\ref{fig: paper prob} illustrates the believed acceptance probability for varying paper qualities.

\subsection{Paper submission process}

We model the paper submission process following the multi-period game outlined in Section~\ref{subsect: multiple submission}. From the multi-period submission model, we know that the researcher will have an optimal submission plan $\mathbf{p^*}(r)=(p^*_0,...,p^*_{n_\text{att}}=1)$, which identifies the optimal journal for each submission attempt $l \in\{1,...,n_\text{att}\}$, based on acceptance probability. When deciding where to submit a paper, the researcher perceives the acceptance probabilities of the journals. We assume that the researcher will only submit a manuscript to a unique journal once. That is, for submission attempt 1, there are $N_J+1$ potential journals ($N_J$ peer-reviewed journals plus the all-acceptance journal), and after $l$ submission attempts, there will be $n_J\equiv (N_J+1)-(l-1)$ journals remaining. For a given submission attempt, for paper $j$, the (perceived) acceptance probabilities is defined as $\mathbf{\hat{p}_j}=(\hat{p}_{j,0},...,\hat{p}_{j,n_J})$. For submission attempt $l$, the researcher would like to submit to journal $k$ that satisfies $\hat{p}_{j,k}=p^*_l$. However, we must recognise that, since our simulation has a finite number of journals, it will not always be possible to find a journal with a (perceived) probability that exactly matches the optimal submission plan. Without loss of generality, assume that the journals are ordered such that $\hat{p}_{j,n_J}\leq...< \hat{p}_{j,0}=1$---recall that journal $k=0$ denotes the all-acceptance journal. For a given $p_l^*$, the researcher will pick journal $k$ that satisfies $\hat{p}_{k+1}<p_l^*\leq \hat{p}_k$ for $k\in\{0,...,n_J-1\}$, or $0<p_l^*\leq \hat{p}_{k}$ for the edge case of $k=n_J$. Note that if the researcher exhausts all possible peer-reviewed journals for a given paper, he/she will have to resort to publishing in the all-acceptance journal. Lastly, we define a time limit in which if a manuscript has not been accepted by a journal after $N_\text{wait}$ years, then the next submission will be to the all-acceptance journal.

\subsection{Peer review process}

When a journal receives an article submission, it will send out three review invitations.%
\footnote{\citet{huisman2017duration} observes that the average number of referee reports per submission is around 2.2 for scientific fields; \citet{fox2017difficulty} also found a similar figure.}
The invited reviewers are randomly chosen from the pool of all researchers (excluding the author).%
\footnote{We note this as a limitation of our model and leave the development of a more sophisticated reviewer selection process to future research.}
Furthermore, a journal will never invite a researcher to review a unique manuscript more than once. After $\eta^{\mathrm{resp}}$ days, the invited researcher will respond accepting or declining the review invitation. If the researcher's review capacity is saturated, he/she will reject the invitation; otherwise, the researcher will accept the review invitation if (and only if)
\begin{equation}
    \hat{q}^J\geq q^R-\theta^R,
\end{equation}
where $\theta_R$ is some tolerance level which we will define shortly.

If the invitation is accepted, the reviewer will take $\eta^{\mathrm{rev}}$ days to return a review score as defined in \eqref{eq: review score}; otherwise, if the invitation is rejected, the journal must select another reviewer (at random) from the reduced pool of researchers. If the pool of researchers is exhausted, then that review slot will default to a review score of $0$. 

Once the journal receives back three review scores, it will accept the paper if (and only if)
\begin{equation}
    \frac{\sum_{i=1}^3\hat{q}^P_{i}}{3} \geq q^J - \theta^J, 
\end{equation}
where $\theta_J$ is some tolerance level which we will define shortly.

Once a paper has been accepted by a journal, we evaluate (i) the quality of the matching, and (ii) the satisfaction of the researcher with the outcome---the researcher will be happier getting published in a better journal. To determine the matching quality, we calculate the quality fit $\delta$ as follows:
\begin{equation}
    \delta_{j,k} = 1-|q^P_j-q^J_k|.
\end{equation}
For the researcher satisfaction, we simply use the utility function defined in \eqref{eq: utility} and calculate $u(1/p_{k^*})$ where $k^*$ is the journal the paper is published in and $p_{k^*}(q^P,q^J_{k^*})$ is the corresponding acceptance probability (given the assumed belief system). Moreover, to account for the associated publication delay, we also calculate a discounted utility
\begin{equation} \label{eq: discounted utility}
    \bar{u}\left(\frac{1}{p_{k^*}};n^*\right)=\beta^{(n^*-1)}u\left(\frac{1}{p_{k^*}}\right),
\end{equation}
where $n^*$ is the number of submission attempts taken.%
\footnote{Note that \eqref{eq: discounted utility} is a crude calculation as it does not consider the fact that the outcome for each submission attempt may be determined at different speeds.}

\section{Deferred acceptance algorithm} \label{sect: DAA}

In this section, we apply DAA matching to our simulation to replace the status quo peer review process. We define a central warehouse responsible for allocating manuscripts to journals. The researchers submit their completed manuscripts to the central warehouse, accompanied with a journal preference list. The warehouse sends the papers out for evaluation, and the journals are given access to the reviews. The journals then submit their preference list (of the manuscripts) to the warehouse. Given the preference lists, the warehouse then matches the papers to the journals using the DAA. We have a many-to-one matching game. This corresponds to the hospital-resident (HR) assignment problem \citep{gale1962,roth1984evolution}. For the HR matching game, there exists a paper- and journal-optimal solution; we will use the journal-optimal solution. We use the Matching library in Python \citep{wilde2020matching} to implement this.

\subsection{Submission and review process}

Papers are submitted to the central warehouse accompanied by a preference list ranking journals from best to worst---this will resemble the researcher's ranking of the journals at the time of submission. Furthermore, assume $0=\hat{q}^J_0<...\leq \hat{q}^J_{N_J}$ such that we can define $\mathbf{\hat{q}^J}=(\hat{q}^J_{N_J},...,\hat{q}^J_0)$ as the preference list of journals---once more, recall that journal $k=0$ is the all-acceptance journal.%
\footnote{It is worth noting that the all-acceptance journal would always be the least-preferred journal.}

Upon receiving a manuscript, the warehouse will send out three review invitations. Like in the status quo, the reviewers are randomly chosen from the pool of all researchers (excluding the author). Unlike in the status quo, now the researcher will reject the invitation only if their review capacity is met. Let us define
\begin{equation}
    \bar{q}_P=\frac{\sum_{i=1}^3\hat{q}_P^i}{3}
\end{equation}
as the aggregate review score of a given paper.%
\footnote{As before, if a review slot is empty and the pool of potential reviewers has been exhausted, the review slot will default to a score of 0.}
Let us define $n_P$ as the number of reviewed, ready-to-allocate papers in the warehouse's inventory and $\mathbf{\bar{q}}_P=(\bar{q}_P^1,...,\bar{q}_P^{n_P})$  as the sorted list of paper scores (best-to-worst). Since the warehouse organises the reviews and ranks the manuscripts centrally, all journals would have the same preference list.%
\footnote{In reality, if the reviews were subjective comments rather than objective scores, the preference lists of the journals can vary.}
When there are enough papers to be assigned to each journal, including the all-acceptance journal ($n_P\geq N_J+1$), the warehouse would solve the corresponding matching game to determine the optimal allocations. For an allocation round, each standard journal is given equal consideration and is designated with a capacity
\begin{equation}
    C_k=\text{Floor}\left(\frac{n_P}{N_J+1}\right), \quad k\in\{1,...,N_J\},
\end{equation}
where any remaining papers will be designated to the all-acceptance journal, such that
\begin{equation}
    C_0=n_P-\sum_{k=1}^{N_J}C_K.
\end{equation}
Once the allocations are determined, each journal publishes the papers it has been assigned. A key feature is that, using DAA, every manuscript will be published in one submission attempt. Once again, we will measure the quality of the matching and satisfaction of the researcher through the quality fit ($\delta$) and utility ($u$ and $\bar{u}$) indicators.

\section{Simulation parameters} \label{sect: parameters}
Now that we have a formal description of the simulation, we need to define its parameters. As much as possible, we based our conclusions on the available literature and common sense.

\subsection{Number of researchers and journals}

\citet{mabe2001growth} and \citet{mabe2003growth} suggest that a growth in the author community of around 100--150 researchers results in the launch of a new journal. This suggests that a journal can accommodate around 100--150 researchers, and a researcher-journal ratio of 125:1 would be an appropriate approximation for the simulation. We decided that 100 (peer-reviewed) journals ($N_J=100$) would provide an appropriate representation of the different journal qualities without having too large a simulation. This will be paired with 12,500 researchers ($N_R=12,500$).

\subsection{Quality and productivity}

Following \citet{lotka1926frequency}, \citet{price1963little}, and \citet{seglen1992}, it is clear that there is a significant skewness of science in different dimensions (including, but not limited to, productivity and paper and journal quality).%
\footnote{For more recent empirical studies, see for example \citet{albarran2011skewness} and \citet{ruiz2014skewness}.}%
\footnote{See also \citet{dorogovtsev2001scaling}, \citet{jackson2007meeting}, and \citet{peterson2010nonuniversal} for theoretical analyses on what may cause the skewness in scientific citedness.}
\citet{shockley1957statistics} suggests that science can be considered as an example of extreme human effort requiring a cumulation of traits. Extreme value distributions are commonly used to model extreme human ability (the best of the best). One such distribution is the Gumbel (double exponential) distribution. The PDF of a generalised right-skewed Gumbel distribution (maximum case) is 
\begin{equation}
    f_{gumbel_R}(x;x_0,\alpha)=\frac{1}{\alpha}\exp{\bigl[-(z_R+e^{-z_R})\bigr]}, 
        \quad z_R=\frac{x-x_0}{\alpha}, \quad x\in(-\infty,\infty),
\end{equation}
where $x_0$ is the location and $\alpha>0$ is the scale. The PDF of a generalised left-skewed Gumbel distribution (minimum case) is
\begin{equation}
     f_{gumbel_L}(x;x_0,\alpha)=\frac{1}{\alpha}\exp{\bigl[-(z_L+e^{-z_L})\bigr]},
        \quad z_L=\frac{x_0-x}{\alpha}, \quad x\in(-\infty,\infty).
\end{equation}
The corresponding CDFs are
\begin{gather}
    F_{gumbel_R} = \exp\bigl[-e^{-z_R}\bigr], \\
    F_{gumbel_L} = 1 - \exp\bigl[-e^{-z_L}\bigr].
\end{gather}

\begin{figure}[t!] 
\centering
\includegraphics[scale=0.7]{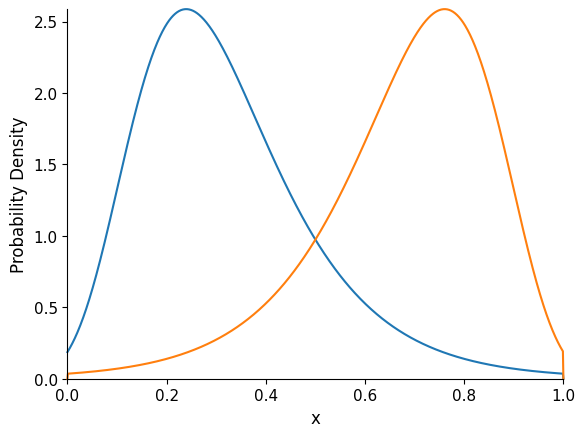}
\caption{Bounded Gumbel distribution with supports $[0,1]$; blue curve is the right-skewed maximum case, and orange curve is the left-skewed minimum case.}
\label{fig: gumbel}
\end{figure}

We define a (99\%) bounded Gumbel
\begin{gather}
    f_{bounded}(x)=\frac{f(x)}{F(b)-F(a)}, \\
    F_{bounded}(x)=\frac{F(x)-F(a)}{F(b)-F(a)},
\end{gather}
with supports $(a,b)$, where $a=F^{-1}(0.005)$ and  $b=F^{-1}(0.995)$. It then follows that if we want a distribution with supports $(x_1,x_2)$, it can be done by calculating
\begin{gather}
    \alpha=\frac{x_2-x_1}{b-a}, \quad x_0=x_1-\alpha a.
\end{gather}
Figure~\ref{fig: gumbel} shows the shape of the bounded Gumbel distributions.%
\footnote{We note that the skewness may not be harsh enough to represent the skewness in scientific metrics, which has been suggested to resemble a power law relationship, approaching linearity in a double-log or semilog plot \citep{lotka1926frequency,seglen1992}. This is another limitation of our model, which we will leave for future research.}

For the journal and researcher qualities, we use the right-skewed distribution bounded to $(0,1)$---this corresponds to a mean of 0.3199. As for researcher writing productivity, \citet{kyvik2003changing} observes during a three-year period (1998-2000) that tenured academic staff averaged three publications per person per year.%
\footnote{Note, however, that this figure also included books and reports.}
However, \citet{RN1} and \citet{kyvik2003changing} suggest that there are around 3 authors per paper. Thus, we would want an average researcher productivity rate of around one paper per year. Since our writing productivity $\eta^w$ represents the number of days it takes to write a paper, we use the left-skewed distribution. We chose supports of (60,500), which corresponds to a mean writing-time of 360 days. For this distribution, we want the writing time to be full days, and thus, we use the ceiling function to get a quasi-discrete distribution.

\subsection{Review duration}

To determine appropriate values for the review and review invitation response times, we will refer to empirical data on first response time (i.e., the duration of the first review round). First response time (FRT) includes the time taken for a journal to initially evaluate a submission and find reviewers, the time reviewers spend reviewing, and the time the editor spends making a final decision (with respect to the review reports) and informing the author(s) \citep{huisman2017duration}. \citet{huisman2017duration} observe that, on average, the FRT is 13 weeks (91 days). 

For our simulation, we assume that the FRT is equal to the review time and review invitation response time:
\begin{equation}
    \mathrm{FRT} = \eta^{\mathrm{resp}} + \eta^{\mathrm{rev}}.
\end{equation}
We decided on an FRT of 90 days, in which it takes a researcher 70 days to review a paper and 20 days to respond to a review invitation ($\eta^{\mathrm{resp}}=20$ and $\eta^{\mathrm{rev}}=70$).

\subsection{Risk aversion, patience, and submission behaviour}

We will use an intertemporal discount factor of $\beta=0.9$, which is a common assumption in economics. As for the (constant level of) relative risk aversion $r$, we will follow estimates from the literature. \citet{groom2019new} estimated that that, in the United Kingdom, $r=1.5$, while \citet{evans2005elasticity} estimated $r$ to be around 1.4 for 20 OECD countries.%
\footnote{Note that relative risk aversion can vary greatly depending on the subpopulation and/or the considered setting; however, these values should provide a good starting range.}
We will use the value of $r=1.5$.

\citet{toroser2017factors} observed that while most papers were accepted within 1--2 submission attempts, some papers can take up to 6 submission attempts to be published. Similarly, \citet{azar2004rejections} estimates that an average paper will take 3--6 submission attempts for publication. Following this, we set $n_\text{att}=6$; that is, on submission attempt 6, the researcher will submit the paper to the all-acceptance journal. Furthermore, we set $N_\text{wait}=35$ years; this figure is a rough approximation of the career length of a researcher.%
\footnote{Nevertheless, it should be recognised that every paper should be published---albeit, in the all-acceptance journal---before this limit is ever reached.}

\subsection{Noise and deviations}

Recall that for our simulation, to calculate $q^P$, $\hat{q}^P$, and $\hat{q}^J$, we have defined noises $\varepsilon^s, \ s\in\{P,\mathrm{rev},J\}$. They account for the variation in quality of papers produced by a researcher and the judgment error the researcher makes when evaluating the quality of a paper or journal. We have defined our quality ratings $q$ to the domain $(0,1)$. Simple Gaussian (normal) noise can push the paper or perceived quality values beyond the specified domain. Thus, we instead use truncated Gaussian noise. The PDF of a truncated normal distribution is given by
\begin{equation}
    f_{\mathrm{truncnorm}}(x;\mu,\sigma,a,b)=\frac{1}{\sigma}
        \frac{\phi\left(\frac{x-\mu}{\sigma}\right)}
        {\Phi\left(\frac{b-\mu}{\sigma}\right)
            -\Phi\Bigl(\frac{a-\mu}{\sigma}\Bigr)}, \quad x\in(a,b),
\end{equation}
where $\phi(\cdot)$ and $\Phi(\cdot)$ are the standard normal PDF and CDF respectively. We sample our noise as follows
\begin{equation}
    \varepsilon^s\sim \text{TruncNorm}(0,\sigma^s,0-q,1-q), \quad s\in\{P,\text{rev}, J\},
\end{equation}
such that $q^P,\hat{q}^P, \hat{q}^J \in(0,1)$. This is equivalent to directly sampling
\begin{align}
    q^P &\sim\text{TruncNorm}(q^R,\sigma^P,0,1), \\
    \hat{q}^P &\sim\text{TruncNorm}(q^P,\sigma^\text{rev},0,1), \\
    \hat{q}^J &\sim\text{TruncNorm}(q^J,\sigma^\text{J},0,1)
\end{align}

For the variability of quality of written papers, we assume moderate fluctuations and set $\sigma^P=0.1$. On the other hand, for the variability in reviews, from the NIPs experiment, it was found that the probability of rejection in the second review, given acceptance in the first review, was around 60\% \citep{langford2015}. Moreover, \citet{bonavia2023noise} observed that, when auditing accepted papers, the probability that a committee of two reviewers disagrees with the publication decision is around 48\%.  This is quite a significant degree of variability, and correspondingly, we assume strong fluctuations and set $\sigma^\text{rev}=0.2$. Lastly, for the journal qualities, we assume that the researchers have access to quantitative measures of journal quality (e.g., SJR, H-index, or impact factor) such that the perception of journal quality only depends on the variation between the different quantitative metrics and subtle subjective preferences. Because of that, we assume weak fluctuations and set $\sigma^J=0.05$.

\subsection{Tolerances}

For the review acceptance and publication tolerances ($\theta^R$, $\theta^J$), we simply assume a 10\% tolerance level. Thus, the review acceptance tolerance for researcher $i$ is 
\begin{equation}
    \theta^R_i=0.1q^R_i,
\end{equation}
and the publication tolerance for journal $k$ is
\begin{equation}
    \theta^J_k=0.1q^J_k.
\end{equation}

\section{Results} \label{sect: results}

We ran the simulation for 100 years (with a rampup of 50 years). Tables~\ref{tab: status quo} and~\ref{tab: daa} show the key metrics from the status quo and DAA simulations, respectively.%
\footnote{The simulations and all results are available in our Github repository (\url{https://github.com/Etienne13/publication_sim}).}
For the status quo simulation, over the 100 years, 1,369,718 papers were written. Of those, 99.05\% were published by the end of the simulation---82.96\% of publications were in a peer-reviewed journal. A total of 11,635,350 reviews were completed, corresponding to an average of 8.58 reviews per publication; 26,345,917 review invitations were sent, corresponding to an average of 19.42 invitations per publication. This translates to a reviewer acceptance rate of 44.16\%---\citet{breuning2015reviewer}, \citet{fox2017difficulty}, and \citet{meyerson2025quantifying} observe reviewer acceptance rates of around 40--60\% corroborating our figure. The mean publication delay was 405.20 days (median: 338 days). Once again, this is a realistic value: \citet{solomon2013} observe an average publication delay of 12.18 months. On average, papers are published after three submission attempts with a quality fit of 0.89. For satisfaction, the mean utility an author receives per publication is 0.43. Considering the publication delay, the mean discounted utility per publication is 0.38. Lastly, the mean quality of papers published in the all-acceptance journal is 0.14.

For the DAA simulation, a total of 1,355,358 papers were written, and 99.74\% were published by the end of the simulation---77.20\% of publications were in a peer-reviewed journal. A total of 4,055,804 reviews were completed, corresponding to an average of 3.00 reviews per publications; 4,146,250 review invitations were sent, corresponding to an average of 3.07 invitations per publication. This translates to a reviewer acceptance rate of 97.82\%. The mean publication delay was 91.24 days (median: 89 days). All papers are published in one submission attempt---as expected---with an average quality fit of 0.89. The mean utility per publication is 0.35; the discounted utility is also 0.35 because every paper is published in one attempt. Lastly, the mean quality of papers published in the all-acceptance journal is 0.17.

It can instantly be noticed that, with DAA matching, the publication process is much more efficient. That is, the amount of review effort required per publication is reduced while the publication process is also sped up. Additionally, the matching quality fit with DAA is almost identical to that of the status quo; that is, the manuscript-journal matching quality is not loss by accelerating the publication process. Furthermore, when the publication delay is accounted for, author satisfaction (discounted utility) under DAA matching is also comparable to the status quo scenario. Lastly, like in the status quo, low-quality papers are being published in the all-acceptance journal.

\begin{table}[ht]
\centering
\footnotesize
\caption{Status quo simulation key metrics.}
\label{tab: status quo}
\renewcommand{\arraystretch}{1.1}
\begin{tabular}{lcccccccccc}
\toprule
\textbf{General} & \multicolumn{2}{c}{\textbf{Count}} & \multicolumn{2}{c}{\textbf{/Papers}} & 
\multicolumn{3}{c}{\textbf{/Publications}}\\
\midrule
Written Papers &
\multicolumn{2}{c}{1,369,718} & \multicolumn{2}{c}{100\%} & 
\multicolumn{3}{c}{---} \\
\midrule
Publications: &
\multicolumn{7}{c}{} \\
\quad \quad \textit{Total} &
\multicolumn{2}{c}{1,356,639} & \multicolumn{2}{c}{99.05\%} & 
\multicolumn{3}{c}{100\%} \\
\quad \quad \textit{Peer-reviewed} &
\multicolumn{2}{c}{1,125,470} & \multicolumn{2}{c}{82.17\%} & 
\multicolumn{3}{c}{82.96\%} \\
\quad \quad \textit{All-acceptance} &
\multicolumn{2}{c}{231,169} & \multicolumn{2}{c}{16.88\%} & 
\multicolumn{3}{c}{17.04\%} \\
\midrule
Reviews &
\multicolumn{2}{c}{11,635,350} & \multicolumn{2}{c}{8.49} & 
\multicolumn{3}{c}{8.58} \\
\midrule
Review Invitations &
\multicolumn{2}{c}{26,345,917} & \multicolumn{2}{c}{19.23} & 
\multicolumn{3}{c}{19.42} \\
\midrule
\textbf{Published papers} & \textbf{Mean} & \textbf{Median} & \textbf{Min} & \textbf{Max} &
\textbf{Q1} & \textbf{Q3} & \textbf{MAD} \\
\midrule
\multirow{2}{*}{Publication delay [days]} 
& 405.20 & 338 & 89 & 1,965 & 218 & 616 & 189 &\\
& (333.86) & (278) & (89)  & (1,965) & (189) & (447) & (129) &\\
\midrule
\multirow{2}{*}{Submission attempts} 
& 3.03 & 2 & 1 & 6 & 2 & 4 & 1 &\\
& (2.42) & (2) & (1)  & (5) & (1) & (3) & (1) &\\
\midrule
\multirow{2}{*}{Quality fit} 
& 0.89 & 0.90 & 0.34 & 1.00 & 0.84 & 0.95 & 0.05 &\\
& (0.90) & (0.91) & (0.34)  & (1.00) & (0.85) & (0.96) & (0.05) &\\
\midrule
\multirow{2}{*}{Utility} 
& 0.43 & 0.46 & 0 & 1.44 & 0.33 & 0.58 & 0.12 &\\
& (0.52) & (0.49) & (0.09)  & (1.44) & (0.40) & (0.60) & (0.10) &\\
\midrule
\multirow{2}{*}{Discounted utility} 
& 0.38 & 0.37 & 0 & 1.42 & 0.27 & 0.51 & 0.12 &\\
& (0.46) & (0.41) & (0.08) & (1.42) & (0.32) & (0.55) & (0.10) &\\
\midrule
All-acceptance paper quality 
& 0.14 & 0.13 & 0.00 & 0.63 & 0.07 & 0.19 & 0.06 & \\
\bottomrule
\end{tabular}
\begin{minipage}{0.9\textwidth}	
				\vspace{0.1cm}
				\footnotesize \textit{NB:} MAD is the median absolute deviation; parenthesised values exclude the all-acceptance journal.
    \end{minipage}
\end{table}

\begin{table}[ht]
\centering
\footnotesize
\caption{DAA simulation key metrics.}
\label{tab: daa}
\renewcommand{\arraystretch}{1.1}
\begin{tabular}{lcccccccccc}
\toprule
\textbf{General } & \multicolumn{2}{c}{\textbf{Count}} & \multicolumn{2}{c}{\textbf{/Papers}} & 
\multicolumn{3}{c}{\textbf{/Publications}}\\
\midrule
Written papers &
\multicolumn{2}{c}{1,355,358} & \multicolumn{2}{c}{100\%} & 
\multicolumn{3}{c}{---} \\
\midrule
Publications: &
\multicolumn{7}{c}{} \\
\quad \quad \textit{Total} &
\multicolumn{2}{c}{1,351,872} & 
\multicolumn{2}{c}{99.74\%} & 
\multicolumn{3}{c}{---} \\
\quad \quad \textit{Peer-reviewed} &
\multicolumn{2}{c}{1,043,700} & 
\multicolumn{2}{c}{77.01\%} & 
\multicolumn{3}{c}{77.20\%} \\
\quad \quad \textit{All-acceptance} &
\multicolumn{2}{c}{308,172} & 
\multicolumn{2}{c}{22.74\%} & 
\multicolumn{3}{c}{22.80\%} \\
\midrule
Reviews &
\multicolumn{2}{c}{4,055,804} & 
\multicolumn{2}{c}{2.99} & 
\multicolumn{3}{c}{3.00} \\
\midrule
Review Invitations &
\multicolumn{2}{c}{4,146,250} & \multicolumn{2}{c}{3.06} & 
\multicolumn{3}{c}{3.07} \\
\midrule
\textbf{Published Papers} & \textbf{Mean} & \textbf{Median} & \textbf{Min} & \textbf{Max} &
\textbf{Q1} & \textbf{Q3} & \textbf{MAD} \\
\midrule
\multirow{2}{*}{Publication delay [days]} 
& 91.24 & 89 & 89 & 176 & 89 & 90 & 0 &\\
& (91.28) & (89) & (89)  & (155) & (89) & (91) & (0) &\\
\midrule
\multirow{1}{*}{Submission attempts} 
& 1 & 1 & 1 & 1 & 1 & 1 & 0 &\\
\midrule
\multirow{2}{*}{Quality fit} 
& 0.89 & 0.90 & 0.30 & 1.00 & 0.84 & 0.95 & 0.06 &\\
& (0.90) & (0.92) & (0.38)  & (1.00) & (0.86) & (0.96) & (0.05) &\\
\midrule
\multirow{2}{*}{(Discounted) Utility} 
& 0.35 & 0.36 & 0 & 1.35 & 0.20 & 0.50 & 0.15 &\\
& (0.45) & (0.42) & (0.04)  & (1.35) & (0.32) & (0.55) & (0.11) &\\
\midrule
All-acceptance paper quality 
& 0.17 & 0.16 & 0.00 & 0.70 & 0.09 & 0.24 & 0.07 & \\
\bottomrule
\end{tabular}
\begin{minipage}{0.9\textwidth}	
				\vspace{0.1cm}
				\footnotesize \textit{NB:} MAD is the median absolute deviation; parenthesised values exclude the all-acceptance journal.	
    \end{minipage}
\end{table}

\subsection{Journal qualities}

\begin{figure}[h!]
    \centering
    \begin{minipage}{0.5\textwidth}
        \centering
        \includegraphics[width=\linewidth]{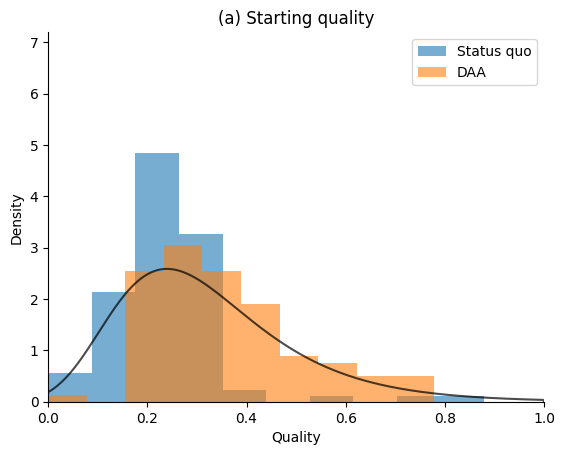}
    \end{minipage}%
    \begin{minipage}{0.5\textwidth}
        \centering
        \includegraphics[width=\linewidth]{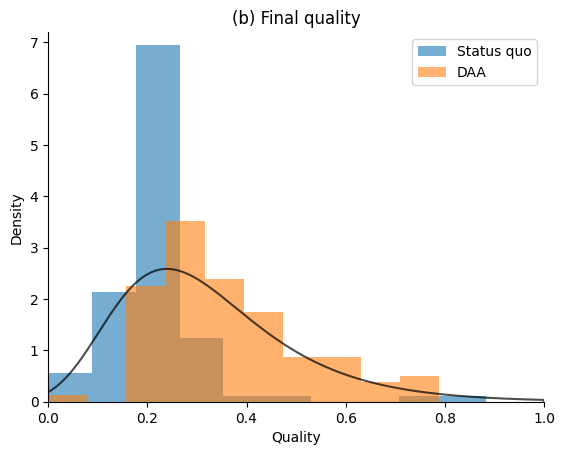}
    \end{minipage}
    \caption{Status quo and DAA journal quality distributions (compared to the underlying PDF).}
    \label{fig: journal quality}
\end{figure}

Another interesting aspect to observe is the journal qualities under the two matching scenarios. Table~\ref{tab: journal quality} outlines the key metrics of the journal quality distributions and Figure~\ref{fig: journal quality} shows the starting and final journal quality distributions. From Table~\ref{tab: journal quality} and Figure~\ref{fig: journal quality}, it can be observed that, after rampup, the journal qualities approach a steady state and do not change majorly, and, more importantly, the journal qualities are noticeably higher with the DAA process (mean final quality: 0.37) than in the status quo (mean final quality: 0.23). Furthermore, under traditional matchmaking (status quo) the journal qualities tend to slightly decrease with the mean quality-change being -3.60\%; in contrast, with DAA matching, the journal qualities tend to slightly increase with the mean quality-change being 1.19\%. These results suggests that, in addition to optimising the peer review process, DAA matchmaking may also be beneficial for journals as it allows them to maintain a higher quality standard.

\begin{table}[ht]
\centering
\footnotesize
\caption{Journal quality distribution summary for the Status quo and DAA simulations.}
\label{tab: journal quality}
\renewcommand{\arraystretch}{1.1}
\begin{tabular}{lccccccccc}
\toprule
 & \textbf{Mean} & \textbf{Median} & \textbf{Min} & \textbf{Max} &
\textbf{Q1} & \textbf{Q3} \\
\midrule
\textit{Status quo:} &&&&&&&\\
\quad \quad Initial Quality
& 0.30 & 0.26 & 0.05 & 0.87 & 0.18 & 0.39 & \\
\quad \quad Starting Quality*
& 0.24 & 0.24 & 0.05 & 0.88 & 0.18 & 0.27 & \\
\quad \quad Final Quality
& 0.23 & 0.23 & 0.05 & 0.88 & 0.18 & 0.26 & \\
\quad \quad Quality Change [\%]**
& -3.60 & -3.97 & -8.72 & 1.01 & -7.01 & 0.00 & \\
\midrule
\textit{DAA:} &&&&&&&\\
\quad \quad Initial Quality
& 0.35 & 0.33 & 0.07 & 0.74 & 0.22 & 0.45 & \\
\quad \quad Starting Quality*
& 0.37 & 0.34 & 0.18 & 0.78 & 0.25 & 0.45 & \\
\quad \quad Final Quality
& 0.37 & 0.34 & 0.19 & 0.79 & 0.26 & 0.45 & \\
\quad \quad Quality Change [\%]**
& 1.19 & 0.85 & -2.39 & 8.66 & -0.17 & 2.21 & \\
\bottomrule
\end{tabular}
\begin{minipage}{0.7\textwidth}	
				\vspace{0.1cm}
				\footnotesize \textit{NB:} These metrics exclude the all-acceptance journal. \\
                *This is the quality at the start of the simulation (after rampup). \\
                **This is the difference between the starting quality and final quality.
    \end{minipage}
\end{table}

\section{Conclusions} \label{sect: conclusion}
Our study shows that a science warehouse that manages reviews and matches papers to journals would be highly beneficial to science. Given the diversity and size of the scientific endeavour, one might wonder how such a science warehouse could be implemented. How could all the researchers and all the journals be encouraged to participate? Who would pay for it? At first sight, these challenges seem insurmountable.

There are several examples, however, that give us hope. We have managed to implement unique identifier systems for researchers (ORCID)\footnote{\url{https://orcid.org}} and publications (DOI)\footnote{\url{https://www.doi.org}}. These are managed and sustained through consortia of publishers and science organisations. The Open Knowledge project provides free and open source software to operate scientific journals, and OpenAlex\footnote{\url{https://openalex.org}} offers an alternative to commercial indexing services such as Scopus. Arxiv\footnote{\url{https://arxiv.org}} and other similar projects offer a free (pre) publishing platform that has seen phenomenal growth. The science community has the expertise to tackle the technical challenges.

The main motivation for researchers and journals is the dramatic increase in efficiency. Authors will be able to publish their articles faster, and journals will no longer have to struggle to find reviewers. But most of all, the DAA rewards honesty. The participants gain no benefit from trying to gamble the system. The algorithm finds the best match for all.

The DAA based peer review process also has some limitations, at least in the initial form we present here. Authors will only have one chance to submit their paper. Currently, authors can use the feedback received from a rejection to improve their work. The DAA peer review process is, therefore, more similar to that of a conference. In the field of computer science, conference proceedings are as recognised as journals, but they operate on a strict schedule. After the main review process, the accepted authors can use the feedback to improve their paper, but rejected papers do not get a second chance. Which brings us to a second limitation.

Low-quality papers are being published in the all-acceptance journal (around 23\%). The authors will still have the opportunity to improve their papers with the feedback received, but they will no longer be able to submit to another journal. The authors might not appreciate this outcome. Still, the proportion of all-acceptance papers mirrors the proportion of papers that do not receive any citations \citep{van_noorden_science_2017}. We might have to acknowledge that the peer review process could have lost its filtering function anyway and only operates as a sorting system \citep{bartneck2010}.

The Delayed Acceptance Algorithm has proven its value in many different application areas, and hence it might not come as a surprise that it can also improve the match-making between papers and journals. Given the great benefits of the algorithm, we hope that our simulation helps promote the idea of using a science warehouse that utilises the Deferred Acceptance Algorithm to better match papers with journals.

\bibliography{references}   

\end{document}